\documentclass{amsart}

\makeatletter
 \def\LaTeX{\leavevmode L\raise.42ex
   \hbox{\kern-.3em\size{\sf@size}{0pt}\selectfont A}\kern-.15em\TeX}
\makeatother

\newcommand{\BibTeX}{{\rm B\kern-.05em{\sc
i\kern-.025emb}\kern-.08em\TeX}}

\newtheorem{thm}{Theorem}[section]
\newtheorem{lem}[thm]{Lemma}

\theoremstyle{definition}
\newtheorem{defn}{Definition}

\newtheorem{rem}{Remark}
\numberwithin{equation}{section}

\begin{document}

\title{Variational splines and Paley-Wiener spaces on combinatorial graphs}

\author{Isaac Pesenson}
\address{Department of Mathematics, Temple University,
Philadelphia, PA 19122} \email{pesenson@math.temple.edu}


 \begin{abstract} Notions of interpolating variational splines and Paley-Wiener spaces
 are introduced on  a combinatorial graph $G$. Both of these definitions explore
 existence of a combinatorial Laplace operator on $G$. The existence and uniqueness of
 interpolating variational spline on a  graph is shown. As an application
of variational splines the paper presents a reconstruction
algorithm  of Paley-Wiener functions on  graphs from their
uniqueness sets.

\end{abstract}

\maketitle

 \section{Introduction and Main Results}

The paper introduces variational splines and Paley-Wiener spaces
 on  combinatorial graphs. Variational splines are defined as
minimizers of Sobolev norms which are introduced in terms of a
combinatorial Laplace operator. It is shown that variational
splines not only interpolate functions but also provide optimal
approximations to them. Paley-Wiener spaces on combinatorial
graphs are defined  by using  spectral resolution of a
combinatorial Laplace operator. The main result of the paper is a
reconstruction algorithm of Paley-Wiener functions from their
uniqueness sets using variational splines.

 The following is a summary of main
notions and results. We consider finite or infinite and in this
case countable connected graphs $G=(V(G),E(G))$,  where $V(G)$ is
its set of vertices and $E(G)$ is its set of edges. We consider
only simple (no loops, no multiple edges) undirected unweighed
graphs. A number of vertices adjacent to a vertex $v$ is called
the degree of $v$ and denoted by $d(v)$. We assume that degrees of
all vertices are bounded from above and we use  notation
$$
d(G)=\max_{v\in V(G)}d(v).
$$
 The space $L_{2}(G)$ is the Hilbert space of all
complex-valued functions $f:V(G)\rightarrow \mathbb{C}$ with the
following inner product
$$
\left<f,g\right>= \sum_{v\in V(G)}f(v)\overline{g(v)}
$$
and the following norm
$$
\|f\|=\|f\|_{0}=\left(\sum_{v\in V(G)}|f(v)|^{2}\right)^{1/2}.
$$

The discrete Laplace operator $\mathcal{L}$ is defined by the
formula  \cite{Ch}
$$
\mathcal{L}f(v)=\frac{1}{\sqrt{d(v)}}\sum_{v\sim
u}\left(\frac{f(v)}{\sqrt{d(v)}}-\frac{f(u)}{\sqrt{d(u)}}\right),
f\in L_{2}(G),
$$
where $v\sim u$ means that $v, u\in V(G)$ are connected by an
edge. It is known that the Laplace operator $\mathcal{L}$ is a
bounded operator in $L_{2}(G)$ which is self-adjoint and positive
definite.  Let $\sigma(\mathcal{L})$ be the spectrum of a
self-adjoint positive definite operator $\mathcal{L}$ in
$L_{2}(G)$, then $\sigma(\mathcal{L})\subset [0,2]$ . In what
follows we will use the notations
$$
\omega_{\min}=\inf_{\omega \in \sigma(\mathcal{L})}\omega,
\omega_{\max}=\sup_{\omega \in \sigma(\mathcal{L})}\omega.
$$

 For a fixed $\varepsilon\geq 0$  the Sobolev
norm is introduced by the following formula
\begin{equation}
\|f\|_{t,\varepsilon}=\left\|(\varepsilon
I+\mathcal{L})^{t/2}f\right\|, t\in \mathbb{R}.\label{Sob}
\end{equation}

The Sobolev space $H_{t,\varepsilon}(G) $ is understood as the
space of functions with the norm (\ref{Sob}). Since the operator
$\mathcal{L}$ is bounded all the spaces $H_{t,\varepsilon}(G),$
coincide as sets.

\bigskip

Variational splines in spaces $L_{2}(\mathbb{R}^{d})$ are
introduced  as functions which minimize certain Sobolev norms
\cite{Schoen}, \cite{D}. Sobolev spaces in $L_{2}(\mathbb{R}^{d})$
can be defined as  domains of powers of the Laplace operator
$\Delta$ in $L_{2}(\mathbb{R}^{d})$ \cite{T}. To construct
variational splines on a graph $G$ we are going to use the same
idea by replacing the classical Laplace operator $\Delta$ by the
combinatorial Laplacian $\mathcal{L}$ in $L_{2}(G)$.

For a given set of indices $I$ (finite or infinite) the notation
$l_{2}$ will be used for the Hilbert space of all sequences of
complex numbers $y=\{y_{i}\}, i\in I$, for which $ \sum_{i\in
I}|y_{i}|^{2}< \infty. $
\bigskip

 \textbf{Variational Problem}

 Given a  subset of vertices $W=\{w\}\subset V(G),  $
 a sequence of complex numbers
$y=\{y_{w}\}\in l_{2}, w\in W$, a positive $t>0$, and an
non-negative $\varepsilon\geq 0$ we consider the following
variational problem:

\bigskip

 \textsl{Find a function  $Y$ from the
space $L_{2}(G)$ which has the following properties:}

1) $Y(w)=y_{w}, w\in W,$

2) $Y$ \textsl{minimizes functional $Y\rightarrow
\left\|(\varepsilon I+\mathcal{L}) ^{t/2}Y\right\|$.}

\begin{rem}
It is convenient to have  such a functional in  the Variational
Problem which is equivalent to  a norm. Thus, if the operator
$\mathcal{L}$ has a bounded inverse in $L_{2}(G)$ (it is a
situation on homogeneous trees of order $q+1, q\geq 2$) then we
will assume that $\varepsilon$ is zero. Otherwise we assume that
$\varepsilon$ is a "small" positive number. In what follows we
will write  $\varepsilon \geq 0$ with understanding that
$\varepsilon = 0$, if the operator $\mathcal{L}$ is invertible in
$L_{2}(G)$ and that $\varepsilon >0$, if $\mathcal{L}$ is not
invertible in $L_{2}(G)$.
\end{rem}

We show that the above variational problem has  a unique solution
$Y_{t,\varepsilon}^{W,y}$. We say that $Y_{t,\varepsilon}^{W,y}$
is a\textit{ variational  spline }of order $t$.  It is also shown
that every spline is a linear combination of fundamental solutions
of the operator $(\varepsilon I+\mathcal{L})^{t}$ and in this
sense it is a polyharmonic function with singularities. Namely it
is shown that every spline satisfies the following equation
$$
(\varepsilon I+\mathcal{L})^{t}Y_{t,\varepsilon}^{W,y}=\sum_{w\in
W}\alpha_{w}\delta_{w},
$$
where $\{\alpha_{w}\}_{w\in
W}=\{\alpha_{w}(Y_{t,\varepsilon}^{W,y})\}_{w\in W}$ is a sequence
from $l_{2}$ and $\delta_{w}$ is the Dirac measure at a vertex
$w\in W$. The set of all such splines for a fixed $W\subset V(G)$
and fixed $t>0, \varepsilon\geq 0,$ will be denoted as
$\mathcal{Y}(W,t,\varepsilon).$

A fundamental solution  $E_{2t,\varepsilon}^{w}, w\in V(G),$ of
the operator $(\varepsilon I+\mathcal{L})^{t} $ is the solution of
the equation

\begin{equation}
(\varepsilon I+\mathcal{L})^{t}E_{2t,\varepsilon}^{w}=\delta_{w},
\end{equation}
where $\delta_{w}$ is the Dirac measure at $w\in V(G)$.

It is shown in the paper that for every  set of vertices
$W=\{w\}, $ every $t>0, \varepsilon \geq 0,$ and for any given
sequence $y=\{y_{w} \}\in l_{2},$ the solution
$Y_{t,\varepsilon}^{W,y}$ of the Variational Problem has a
representation
$$
Y_{t,\varepsilon}^{W,y}=\sum_{w\in W} y_{w}L^{W,w}_{t,
\varepsilon},
$$
where $L^{W,w}_{t, \varepsilon}$ is the so called Lagrangian
spline, i.e. it is a solution of the same Variational Problem with
constraints $ L^{W,w}_{t, \varepsilon}(v)=\delta_{w,v}, w\in W,$
where $\delta_{w,v}$ is the Kronecker delta. Another
representation is
$$
Y_{t,\varepsilon}^{W,y}=\sum_{w\in
W}\alpha_{w}(Y_{t,\varepsilon}^{W,y})E^{w}_{2t,\varepsilon},\label{fund.sol.representation}
$$
where $\{\alpha_{w}(Y_{t,\varepsilon}^{W,y})\}_{w\in W}$ is a
sequence in $l_{2}$.

Given a function $f\in L_{2}(G)$ we will say that the spline
$Y_{t,\varepsilon}^{W,f}$  interpolates $f$ on $W$ if
$Y_{t,\varepsilon}^{W,f}(w)=f(w)$ for all $w\in W$.
 It is shown in the Theorem 2.4 that for a
given function $f\in L_{2}(G)$ its interpolating spline $Y_{t,
\varepsilon}^{W,f}$ is always an optimal approximation (modulo
given information).

\begin{rem}

It is important to realize that for a fixed set $W\subset V(G)$
and fixed $t, \varepsilon\geq 0,$ the correspondence
\begin{equation}
\{y_{w}\}\rightarrow
\left\{\alpha_{w}(Y_{t,\varepsilon}^{W,y})\right\}, y=\{y_{w}\}\in
l_{2},
\end{equation}
where $Y^{W,y}_{t, \varepsilon}$ is a spline, depends just on the
geometry of $G$ and $W$. In other words the map (1.3) is
responsible for the connection between "analysis" on $G$ and its
geometry.

\end{rem}

Our main goal is to develop spline interpolation and approximation
in the so-called Paley-Wiener spaces.

 Paley-Wiener spaces on
$\mathbb{R}^{d}$ are denoted $PW_{\omega}(\mathbb{R}), \omega>0,$
and contain functions $f\in L_{2}(\mathbb{R})$ whose
$L_{2}$-Fourier transform
$$
\hat{f}(\xi)=\frac{1}{\sqrt{2\pi}}\int_{-\infty}^{+\infty}
f(x)e^{- i x \xi}dx
$$
has support in $[-\omega,\omega]$. The classical sampling theorem
says  that if $f\in PW_{\omega}(\mathbb{R})$ then $f$ is
completely determined by its values at points $k\pi/\omega, k\in
\mathbb{Z}$, and can be reconstructed in a stable way from the
samples $f(k\pi/\omega)$ by using the so-called cardinal series
$$
 f(x)= \sum_{k\in \mathbb{Z}} f\left(\frac{k\pi}{\omega}\right)\frac{\sin(\omega
x-k\pi)}{\omega x-k\pi },
$$
where convergence is understood in the  $L_{2}$-sense. In papers
\cite{S1}, \cite{S2},  \cite{Gol}, \cite{LMad}, \cite{Pes99},
\cite{Pes30}, splines were used as a tool for reconstruction of
Paley-Wiener
 functions from their uniqueness sets.

The  Paley-Wiener spaces and in particular  sampling problems in
these spaces attracted attention of many mathematicians
\cite{B64}, \cite{BM67}, \cite{DS}, \cite{Lan67}, \cite{PW34},
\cite{OS}, \cite{LS02}. C. Shannon \cite {SW}, \cite{S}, suggested
to use the sampling theory in Paley-Wiener spaces as a theoretical
foundation for practical problems in signal analysis and
information theory. Since then the sampling theory found many
other applications in particular in image reconstruction and
learning theory \cite{SS1}, \cite {SS2}.
 Some of the ideas and methods of the sampling theory of Paley-Wiener
 functions  were recently
extended to the cases of Riemannian manifolds, groups, and quantum
graphs \cite{EEKK1}, \cite{EEKK2},  \cite{FP1}, \cite{FP2},
\cite{F}, \cite{FG}, \cite{Pes98}- \cite{P07}. Splines on
manifolds and quantum graphs were developed in \cite{P1},
\cite{P2}.

 To define Paley-Wiener spaces on combinatorial graphs we use the fact that
 the Laplace operator $\mathcal{L}$ is a self-adjoint positive definite
 operator in the Hilbert space $L_{2}(G)$. According to the spectral
theory \cite{BS} there exist a direct integral of Hilbert spaces
$X=\int X(\lambda)dm (\lambda )$ and a unitary operator $F$ from
$L_{2}(G)$ onto $X$, which transforms domain of $\mathcal{L}^{s},
s\geq 0,$ onto $X_{s}=\{x\in X|\lambda^{s}x\in X \}$ with norm
$$
\|x(\tau )\|_{X_{s}}= \left (\int_{\sigma(\mathcal{L})}
\lambda^{2s} \|x(\lambda )\|^{2}_{X(\lambda)} d m(\lambda) \right
)^{1/2}
$$
and $F(\mathcal{L}^{s} f)=\lambda ^{s} (Ff)$.
 We introduce the following notion of discrete Paley-Wiener spaces.
\begin{defn}
Given an $\omega\geq 0$ we will say that a function $f$ from
$L_{2}(G)$ belongs to the Paley-Wiener space $PW_{\omega}(G)$ if
its "Fourier transform" $Ff$ has support in
 $[0, \omega ] $.
\end{defn}

To be more consistent with the definition of the classical
Paley-Wiener spaces we should consider the interval
$[0,\omega^{2}]$ instead of $[0,\omega]$. We prefer our choice
because it makes formulas and notations simpler.

Since the operator $\mathcal{L}$ is bounded every function from
$L_{2}(G)$ belongs to a certain  Paley-Wiener space
$PW_{\omega}(G)$ for some $\omega\in \sigma(\mathcal{L})$ and we
have the following stratification
$$
L_{2}(G)=PW_{\omega_{\max}}(G)=\bigcup_{ \omega\in
\sigma(\mathcal{L})}PW_{\omega}(G), PW_{\omega_{1}}(G)\subseteq
PW_{\omega_{2}}(G), \omega_{1}<\omega_{2}.
$$

Different properties of the spaces $PW_{\omega}(G)$ and in
particular a generalization of the Paley-Wiener Theorem are
collected in the Theorem 3.1.

For a subset $S\subset V(G)$ (finite or infinite) the notation
$L_{2}(S)$ will denote the space of all functions from $L_{2}(G)$
with support in $S$:
$$
L_{2}(S)=\{\varphi\in L_{2}(G), \varphi(v)=0, v\in V(G)\backslash
S\}.
$$

\begin{defn}
We say that a set of vertices $U\subset V(G)$ is a uniqueness set
for a space  $PW_{\omega}(G), \omega>0,$ if for any two functions
from $PW_{\omega}(G)$ the fact that they coincide on $U$ implies
that they coincide on $V(G)$.
\end{defn}
\begin{defn}
We say that a set of vertices $S\subset V(G)$ is a $\Lambda$-set
if for any $\varphi\in L_{2}(S)$ it admits a Poincare  inequality
with a constant $\Lambda>0$
$$
\|\varphi\|\leq \Lambda\|\mathcal{L}\varphi\|, \varphi\in
L_{2}(S), \Lambda>0.
$$
The infimum of all $\Lambda>0$ for which $S$ is a $\Lambda$-set
will be called the Poincare constant of the set $S$ and denoted by
$\Lambda(S)$.
\end{defn}

It is shown in the Theorem 3.4 that if a set $S\subset V(G)$  is a
$\Lambda$-set,  then its complement   $U=V(G)\backslash S$ is a
uniqueness set for any space $PW_{\omega}(G)$ with $ \omega<
1/\Lambda$.
 Since $L_{2}(G)=PW_{\omega_{\max}}(G)$ \textit{every} function in
 $L_{2}(G)$ belongs to a certain Paley-Wiener space and
one cannot expect that non-trivial uniqueness sets there exist for
functions from \textit{every} Paley-Wiener subspace.  But it is
reasonable to expect that uniqueness sets exist for Paley-Wiener
spaces $PW_{\omega}(G)$ with relatively small $\omega>0$. It will
be shown (see Section 3)  that for every graph $G$ there exists a
constant $\Omega_{G}\geq 1$ such that for $0<\omega<\Omega_{G}$
functions from $PW_{\omega}(G)$  can be determined by using their
values only on certain subsets of vertices. Namely, it is shown
that for any graph $G$ spaces $PW_{\omega}(G)$ with
$$
0<\omega<\sqrt{1+\frac{1}{d(G)}}=\Omega_{G}>1
$$
have non-trivial uniqueness sets. A more detailed description of
uniqueness sets will be given in a separate paper.

The main result of the present article is obtained in Section 4
and can be stated in the following form.

\begin{thm}

1) Assume that $\mathcal{L}$  is invertible in $L_{2}(G)$ . If $S$
is a $\Lambda$-set then any $f\in PW_{\omega}(G)$ with $
\omega<1/\Lambda$ can be reconstructed from its values on
$U=V(G)\setminus S$ as the following limit
$$
f=\lim_{k\rightarrow \infty}Y_{k}^{U, f},k=2^{l}, l\in \mathbb{N},
$$
where $Y_{k}^{U, f}$ is a spline interpolating $f$ on the set
$U=V(G)\setminus S$ and the error estimate is
$$
\left\|f-Y_{k}^{U, f}\right\|\leq 2\gamma^{k}\left\|f\right\|,
\gamma=\Lambda\omega<1, k=2^{l}, l\in \mathbb{N}.
$$

2) If the operator $\mathcal{L}$ in $L_{2}(G)$ is not invertible,
then for any $\Lambda$-set $S$ and any $0<\varepsilon <1/\Lambda,$
every function $f\in PW_{\omega}(G),$ where
$$
0<\omega<\frac{1}{\Lambda}-\varepsilon,
$$
 can be reconstructed from its
values on $U=V(G)\setminus S$ as the following limit
$$
f=\lim_{k\rightarrow \infty}Y_{k,\varepsilon}^{U, f},k=2^{l}, l\in
\mathbb{N},
$$
where $Y_{k,\varepsilon}^{U, f}$ is a spline interpolating $f$ on
the set $U=V(G)\setminus S$ and the error estimate is given by
$$
\left\|f-Y_{k,\varepsilon}^{U, f}\right\|\leq 2\gamma^{k}\|f\|,
\gamma=\Lambda(\omega+\varepsilon)<1,k=2^{l}, l\in \mathbb{N}.
$$
\end{thm}

We know two papers \cite{FrT}, \cite{G},  in which authors
consider sampling on $\mathbb{Z}^{n}$ and $\mathbb{Z}_{N}$ and one
paper \cite{MSW} were sampling on $\mathbb{Z}^{n}$ was used to
prove some deep results in discrete harmonic analysis. But our
approach to the problem and our results are very different from
the methods and results of these papers.

\section{Variational splines on combinatorial graphs}

We are going to use the same notations and the same Variational
Problem which were defined  in the Introduction.

\begin{thm}
For  every  set of vertices $W=\{w\}, $ all $k>0,\varepsilon\geq
0,$ and for any given sequence $y=\{y_{w} \}\in l_{2},$ the
Variational Problem has a unique solution.
\end{thm}
\begin{proof}

Consider the set $\mathcal{M}_{0}(W)\subset L_{2}(G), $ of all
functions from $L_{2}(G)$ whose restriction to $W=\{w\}$ is zero.
This  is a closed subspace of $L_{2}(W)$.

Given a sequence of complex numbers $y=\{y_{w}\}\in l_{2},$ the
 linear manifold $\mathcal{M}(W,y)$
 of all functions $f$ from
$L_{2}(G)$ such that $f(w)=y_{w}$ is a shift of the closed
subspace $\mathcal{M}_{0}(W)$, i.e.
\begin{equation}
\mathcal{M}(W,y)=\mathcal{M}_{0}(W)+g,\label{space}
\end{equation}
 where $g$ is
any function from $L_{2}(G)$ such that for all $w \in W$ one has
$g(w)=y_{w}.$

Consider the orthogonal projection $h_{t,\varepsilon}$ of the
function $g\in L_{2}(G)$ from (\ref{space}) onto the space
$\mathcal{M}_{0}(W)$ with respect to the inner product in
$H_{t,\varepsilon}(G), t>0$:

$$\left<f_{1},
f_{2}\right>_{H_{t,\varepsilon}(G)}= \left<(\varepsilon
I+\mathcal{L})^{t/2}f_{1},(\varepsilon
I+\mathcal{L})^{t/2}f_{2}\right>_{L_{2}(G)}.
$$

The function  $Y_{t,\varepsilon}^{W,y}=g-h_{t,\varepsilon}$ is the
solution to the above variational problem. Indeed, it is clear
that $Y_{t,\varepsilon}^{W,y}\in \mathcal{M}(W,y)$. To show that
$Y_{t,\varepsilon}^{W,y}$ minimizes the functional
$$Y\rightarrow \|(\varepsilon I+\mathcal{L})^{t/2}Y\|$$
on the set $\mathcal{M}(W,y)$ we note that any function from
$\mathcal{M}(W,y)$ can be written in the form
$Y_{t,\varepsilon}^{W,y}+\psi,$ where $\psi\in
\mathcal{M}_{0}(W)$. Since
$Y_{t,\varepsilon}^{W,y}=g-h_{t,\varepsilon}$ is orthogonal to
$\mathcal{M}_{0}(W)$ in $H_{t,\varepsilon}(G)$ we obtain for any
$\sigma\in \mathbb{C}$
$$
\|(\varepsilon I+\mathcal{L})^{t/2}(Y_{t,\varepsilon}^{W,y}+\sigma
\psi)\|^{2}= \|(\varepsilon
I+\mathcal{L})^{t/2}Y_{t,\varepsilon}^{W, y}\|^{2}+
|\sigma|^{2}\|(\varepsilon I+\mathcal{L})^{t/2}\psi\|^{2}, \psi\in
\mathcal{M}_{0}(W),
$$
that means that the function $Y_{t,\varepsilon}^{W,y}$ is the
minimizer.

The fact that the minimizer is unique follows from the well-known
properties of Hilbert spaces. The proof is complete.
\end{proof}

The following result shows that every solution of the Variational
Problem 1)-2) should be  a "polyharmonic function" with
"singularities" on the set $W$.

\begin{thm}
For every  set of vertices  $W=\{w\}, w\in V(G), $ every
$t>0,\varepsilon\geq 0,$ and for any given sequence $y=\{y_{w}
\}\in l_{2},$ the solution $Y_{t,\varepsilon}^{W,y}$ of the
Variational Problem
 satisfies the following equation
\begin{equation}
(\varepsilon I+\mathcal{L})^{t}Y_{t,\varepsilon}^{W,y}=\sum_{w\in
W}\alpha_{w}\delta_{w},\label{distrib. equation}
\end{equation}
where $\{\alpha_{w}\}_{w\in
W}=\{\alpha_{w}(Y_{t,\varepsilon}^{W,y})\}_{w\in W}$ is a sequence
from $l_{2}$. Conversely, if a function satisfies equation
(\ref{distrib. equation}) then it is a spline.
\end{thm}
\begin{proof}

If $\delta_{w}$ is a Dirac function concentrated  at a point $w\in
W$ then for any $\phi\in L_{2}(G)$ the function

$$\psi=\phi-\sum_{w\in W}\phi(w)\delta_{w}
$$
belongs to $\mathcal{M}_{0}(W)$ and because every solution of the
above Variational Problem 1)-2) is orthogonal to
$\mathcal{M}_{0}(W)$ in the Hilbert space $H_{2t,\varepsilon}(G)$
we obtain

$$
0=\sum_{v\in V(G)}\left(\varepsilon
I+\mathcal{L}\right)^{t/2}Y_{t,\varepsilon}^{W,y}(v)\overline{
\left(\varepsilon I+\mathcal{L}\right)^{t/2}\psi(v)}.
$$

It implies that
$$\sum_{v \in V(G)}\left(\varepsilon
I+\mathcal{L}\right)^{t}Y_{t,\varepsilon}^{W,y}(v)\overline{\phi(v)}=
\sum_{w\in
W}\left<Y_{t,\varepsilon}^{W,y},\delta_{w}\right>_{H_{t,\varepsilon}(G)}\overline{\phi(w)},
\phi\in L_{2}(G).
$$

 In other words
$\left(\varepsilon
I+\mathcal{L}\right)^{t}Y_{t,\varepsilon}^{W,y}$ is a function of
the form

$$
(\varepsilon I+\mathcal{L})^{k}Y_{t,\varepsilon}^{W,y}=\sum_{w\in
W}\alpha_{w}(Y_{t,\varepsilon}^{W,y})\delta_{w},
$$
where
$\alpha_{w}(Y_{t,\varepsilon}^{W,y})=\left<Y_{t,\varepsilon}^{W,y},
\delta_{w}\right>_{H_{t,\varepsilon}(G)} \in l_{2}.$ Thus we
proved that every solution of the Variational Problem is a
solution of (\ref{distrib. equation}).  The converse is obvious.
The Theorem is proved.
\end{proof}

A fundamental solution  $E_{2t,\varepsilon}^{v}, v\in V(G)$,  of
the operator $(\varepsilon I+\mathcal{L})^{t}, $ is a solution of
the equation
\begin{equation}
(\varepsilon I+\mathcal{L})^{t}E_{2t,\varepsilon}^{v}=\delta_{v},
\end{equation}
where $\delta_{v}$ is the Dirac measure at $v\in V(G)$. The
following Theorem explains the structure of splines and it follows
from (\ref{distrib. equation}) and linearity of the set of splines
which is a consequence of the last Theorem.

\begin{thm}
For every  set of vertices  $W=\{w\}, $ every $t>0, \varepsilon
\geq 0,$ and for any given sequence $y=\{y_{w} \}\in l_{2},$ the
solution $Y_{t,\varepsilon}^{W,y}$ of the Variational Problem has
a representation
$$
Y_{t,\varepsilon}^{W,y}=\sum_{w\in W} y_{w}L^{W, w}_{t,
\varepsilon},
$$
where $L^{W, w}_{t, \varepsilon}$ is the so called Lagrangian
spline, i.e. it is a solution of the same Variational Problem with
constrains $ L^{W, w}_{t, \varepsilon}(v)=\delta_{w,v}, w\in W,$
where $\delta_{w,v}$ is the Kronecker delta. Another
representation is
\begin{equation}
Y_{t,\varepsilon}^{W,y}=\sum_{w\in
W}\alpha_{w}(Y_{t,\varepsilon}^{W,y})E^{w}_{2t,\varepsilon},\label{fund.sol.representation}
\end{equation}
where $\{\alpha_{w}(Y_{t,\varepsilon}^{W,y})\}_{w\in W}$ is a
sequence in $l_{2}$.
\end{thm}

Now we are going to show that variational interpolating splines
provide an optimal approximation.

\begin{defn}
For the given  $W\subset V(G), f\in L_{2}(G), t>0, \varepsilon\geq
0, K>0,$ the notation $Q(W,f,t, \varepsilon, K)$ will be used for
a set of all functions $g$ in $L_{2}(G)$ such that

1) $g(w)=f(w), w\in W,$

and

2) $\left\|(\varepsilon I+\mathcal{L})^{t/2}g\right\|\leq K.$

\end{defn}

 It is easy to verify that every set $Q(W,f,k,
\varepsilon, K)$ is convex, bounded, and closed. The next Theorem
shows that for a given function $f\in L_{2}(G)$ its interpolating
spline $Y_{t, \varepsilon}^{W,f}$ is always an optimal
approximation (modulo given information).
\begin{thm}
The following statements hold true:

1) If $ K<\left\|(\varepsilon I+\mathcal{L})^{t/2}Y_{t,
\varepsilon}^{W,f}\right\|$ then the set $Q(W,f,t, \varepsilon,
K)$ is empty.

2) Every variational spline $Y_{t, \varepsilon}^{W,f}$ is the
center of the convex set $Q(W,f,t, \varepsilon, K)$.  As a result
the following inequalities holds true for any $g\in
Q(W,f,t,\varepsilon, K)$
$$\left\|Y_{t, \varepsilon}^{W,f}-g\right\|_{H_{t, \varepsilon}(G)}
   \leq \frac{1}{2} diam
   Q(W,f,t, \varepsilon, K),
   $$
 and
$$
  \left\|Y_{t, \varepsilon}^{W,f}-g\right\|_{L_{2}(G)}
  \leq \frac{1}{2}\left\|(\varepsilon I+\mathcal{L})^{-t/2}\right\| diam Q(W,f,t,\varepsilon,
  K),
$$
where $diam$ is taken with respect to the  norm  of the Sobolev
space $H_{t, \varepsilon}(G)$.
\end{thm}
\begin{proof}
Given a function $f\in L_{2}(G)$ the
 linear manifold $\mathcal{I}(W,f)$
 is the set of all functions $g$ from
$L_{2}(G)$ such that $f(w)=g(w), w\in W$. Let us note that the
distance from zero to the subspace $\mathcal{I}(W, f), $ in the
metric of the space $H_{t,\varepsilon}(G)$ is exactly the Sobolev
norm of the unique spline $Y_{k,\varepsilon}^{W,f}\in
\mathcal{I}(W, f).$
 This norm can be expressed in
terms of the sequence $(Y_{t, \varepsilon}^{W,f})(w)=f(w), w\in
W,$ and the sequence $\{\alpha_{w}(Y_{t,\varepsilon}^{W,f})\},
w\in W,$ from the representation
$$
Y_{t,\varepsilon}^{W,f}=\sum_{w\in
W}\alpha_{w}(Y_{t,\varepsilon}^{W,f})E^{w}_{2t,\varepsilon}.
$$
Indeed,

$$
\|Y_{t,\varepsilon}^{W,f}\|_{H_{t,\varepsilon}(G)}=\left<(\varepsilon
I+\mathcal{L})^{t/2}Y_{t, \varepsilon}^{W,f}, (\varepsilon
I+\mathcal{L})^{t/2}Y_{t,\varepsilon}^{W,f}\right
>^{1/2}=
$$
$$
\left<(\varepsilon I+\mathcal{L})^{t}Y_{t,
\varepsilon}^{W,f},Y_{t, \varepsilon}^{W,f}\right>^{1/2}=
\left<\sum_{w\in W}\alpha_{w}(Y_{t,
\varepsilon}^{W,f})\delta_{w},Y_{t,
\varepsilon}^{W,f}\right>^{1/2}=
$$
$$
\left(\sum_{w\in W}\alpha_{w}(Y_{t,
\varepsilon}^{W,f})f(w)\right)^{1/2}.
$$
 It shows that  the intersection
$$
Q(W,f,t,\varepsilon, K)=\mathcal{I}(W, f)\bigcap B_{t,
\varepsilon}(0, K),
$$
where $B_{t, \varepsilon}(0, K)$ is the ball in
$H_{t,\varepsilon}(G)$ whose center is zero and the  radius is
$K$, is not empty if and only if

$$
K\geq\left\|Y_{t, \varepsilon}^{W,f}\right\|_{H_{t,
\varepsilon}(G)}=\left(\sum_{w\in W}\alpha_{w}(Y_{t,
\varepsilon}^{W,f})f(w))\right)^{1/2}.
$$
The first part of the Theorem is proved.

Now  we are going to show that for a given function $f$ the
interpolating spline $Y_{t, \varepsilon}^{W,f}$ is the center of
the convex, closed and bounded set $Q(W,f,t, \varepsilon, K)$ for
any $K\geq \|Y_{t,\varepsilon}^{W,f}\|_{H_{t, \varepsilon}(G)}$ .
In other words it is sufficient to show that if
$$
Y_{t, \varepsilon}^{W,f}+h\in Q(W,f,t, \varepsilon, K)
$$
for some function $h$ from the Sobolev space $H_{t,
\varepsilon}(G)$ then the function $Y_{t, \varepsilon}^{W,f}-h$
also belongs to the same intersection.
 Indeed, since $h$ is zero on the set $W$ one has

$$
\left <(\varepsilon I+ \mathcal{L})^{t/2}Y_{t, \varepsilon}^{W,f},
(\varepsilon I+\mathcal{L})^{t/2}h \right>=\left <(\varepsilon I+
\mathcal{L})^{t}Y_{t, \varepsilon}^{W,f}, h \right>=0.
$$ But then

$$\left\|(\varepsilon I+\mathcal{L})^{t/2}(Y_{t, \varepsilon}^{W,f}+h)\right\|_{L_{2}(G)}=
\left\|(\varepsilon I+\mathcal{L})^{t/2}\left(Y_{t,
\varepsilon}^{W,f}-h\right)\right\|_{L_{2}(G)}.
$$
In other words,
 $$\left\|(\varepsilon I+\mathcal{L})^{t/2}(Y_{t, \varepsilon}^{W,f}-h)\right\|_{L_{2}(G)}\leq K $$
 and because $Y_{t, \varepsilon}^{W,f}+h$ and $Y_{t, \varepsilon}^{W,f}-h$ take the same
values on $W$ the function $Y_{t, \varepsilon}^{W,f}-h$ belongs to
 $Q(W,f,t, \varepsilon, K).$
 It is clear that  the following inequality
holds true
$$
  \|Y_{t, \varepsilon}^{W,f}-g\|_{H_{t, \varepsilon}(G)}\leq \frac{1}{2} diam  Q(W,f,t,\varepsilon, K)
$$
for any $g\in Q(W,f,t,\varepsilon, K)$. Using this inequality one
obtains

$$\left\|Y_{t, \varepsilon}^{W,f}-g\right\|_{L_{2}(G)}=
   \left \|(\varepsilon I+\mathcal{L})^{-t/2}(\varepsilon I+\mathcal{L})^{t/2}\left(Y_{t, \varepsilon}^{W,f}-g\right)\right
   \|_{L_{2}(G)}
   $$
   $$
  \leq \frac{1}{2}\left\|(\varepsilon I+\mathcal{L})^{-t/2}\right\| diam Q(W,f,t,\varepsilon, K).
   $$

The Theorem is proven.
\end{proof}
\bigskip

\section{Paley-Wiener spaces on combinatorial graphs}

The Paley-Wiener spaces $PW_{\omega}(G), \omega>0,$ were
introduced in the Definition 1 of the Introduction. Since the
operator $\mathcal{L}$ is bounded it is clear that every function
from $L_{2}(G)$ belongs to a certain Paley-Wiener space.  Note
that if
$$
\omega_{\min}=\inf _{\omega\in \sigma(\mathcal{L})}\omega
$$
then the space $PW_{\omega}(G)$ is not trivial if and only if
$\omega\geq \omega_{\min}$.

Using the spectral resolution of identity $P_{\lambda}$ we define
the unitary group of operators by the formula
$$
e^{it\mathcal{L}}f=\int_{\sigma(\mathcal{L})}e^{it\tau}dP_{\tau}f,
f\in L_{2}(G), t\in \mathbb{R}.
$$

 The next theorem can be considered
as a form of the Paley-Wiener theorem and it essentially follows
from a more general result in \cite{Pes00}.
\begin{thm}
The following statements hold true:

1) $f\in PW_{\omega}(G)$ if and only if for all $s\in
\mathbb{R}_{+}$ the following Bernstein inequality takes place
\begin{equation}
\|\mathcal{L}^{s}f\|\leq \omega^{s}\|f\|;
\end{equation}

2) the norm of the operator $\mathcal{L}$ in the space
$PW_{\omega}(G)$ is exactly $\omega$;

3) $f\in PW_{\omega}(G)$ if and only if the following holds true
$$
\lim_{s\rightarrow \infty} \|\mathcal{L}^{s}f\|^{1/s}=\omega, s\in
\mathbb{R}_{+};
$$

 4)  $f\in PW_{\omega}(G)$ if and only if for every $g\in
L_{2}(G)$ the scalar-valued function of the real variable $t\in
\mathbb{R}^{1}$
$$
\left<e^{it\mathcal{L}}f,g\right>=\sum_{v\in V}
e^{it\mathcal{L}}f(v)\overline{g(v)}
$$
 is bounded on the real line and has an extension to the complex
plane as an entire function of the exponential type $\omega$;

5) $f\in PW_{\omega}(G)$ if and only if the abstract-valued
function $e^{it\mathcal{L}}f$  is bounded on the real line and has
an extension to the complex plane as an entire function of the
exponential type $\omega$;

6) $f\in PW_{\omega}(G)$ if and only if the solution $u(t,v), t\in
\mathbb{R}^{1}, v\in V(G),$ of the Cauchy problem for the
corresponding Schrodinger equation
$$
i\frac{\partial u(t,v)}{\partial t}=\mathcal{L}u(t,v),
u(0,v)=f(v), i=\sqrt{-1},
$$
 has analytic extension $u(z,v)$ to the
complex plane $\mathbb{C}$ as an entire function and satisfies the
estimate
$$
\|u(z, \cdot)\|_{L_{2}(G)}\leq e^{\omega|\Im z|}\|f\|_{L_{2}(G)}.
$$

\end{thm}

 We prove here only the first part of the Theorem.

\begin{lem}A function $f\in L_{2}(G)$ belongs to $PW_{\omega}(G)$ if and only if the
following Bernstein inequality holds true for all  $s\in
\mathbb{R}_{+}$
\begin{equation}
\|\mathcal{L}^{s}f\|\leq\omega^{s}\|f\|. \label{Berns3}
\end{equation}\label{Thm2}
\end{lem}
\begin{proof}We use the spectral theorem for the operator
$\mathcal{L}$ in the space $L_{2}(G)$ in the form it was presented
in the Introduction.

 Let $f$ belongs to the space $PW_{\omega }(G)$ and $\mathcal{F}_{\mathcal{L}}f=x\in
X$. Then

$$
\left(\int ^{\infty}_{0}\lambda
^{2s}\|x(\lambda)\|^{2}_{X(\lambda)}dm(\lambda )\right)^{1/2}=
\left(\int^{\omega}_{0}\lambda^{2s}\|x(\lambda
)\|^{2}_{X(\lambda)}dm(\lambda)\right)^{1/2} \leq \omega
^{s}\|x\|_{X} , s\in \mathbb{R}_{+} ,
$$
which gives Bernstein
inequality for $f$.

Conversely, if $f$ satisfies Bernstein inequality then
$x=\mathcal{F}_{\mathcal{L}}f$ satisfies $\|x\|_{X_{s}} \leq
\omega^{s}\|x\|_{X}.$ Suppose that there exists a set $\sigma
\subset [0, \infty )\setminus [ 0, \omega]$ whose $m $-measure is
not zero and $x|_{\sigma }\neq 0.$ We can assume that $\sigma
\subset  [\omega +\epsilon , \infty )$ for some $\epsilon
>0.$ Then for any $s\in \mathbb{R}_{+}$ we have

$$
\int _{\sigma }\|x(\lambda )\|^{2}_{X(\lambda)}dm (\lambda ) \leq
 \int ^{\infty
}_{\omega +\epsilon}\lambda ^{-2s}\| \lambda
^{s}x(\lambda)\|^{2}_{ X(\lambda)}d\mu
\leq\|x\|^{2}_{X}\left(\omega /\omega +\epsilon \right)^{2s},
$$
which shows that or $x(\lambda)$ is zero on $\sigma $ or $\sigma $
has measure zero.
\end{proof}

 The  Theorem 3.1  shows that
the notion of Paley-Wiener functions of type $\omega$ on a
combinatorial graph can be completely understood in terms of
familiar  entire functions of exponential type $\omega$ bounded on
the real line.

The notion of $\Lambda$-sets was introduced in the Definition 3 in
the Introduction. The role of $\Lambda$-sets is explained in the
following Theorem.

\begin{thm}
If a set $S\subset V(G)$  is a $\Lambda$-set,  then  the set
$U=V(G)\backslash S$ is a uniqueness set for any space
$PW_{\omega}(G)$ with $ \omega< 1/\Lambda$.
\end{thm}

\begin{proof}
If $f,g\in PW_{\omega}(G)$ then $f-g\in PW_{\omega}(G)$ and
according to the Theorem 3.1 the following Bernstein inequality
holds true
\begin{equation}
\|\mathcal{L}(f-g)\|\leq \omega\|f-g\|.
\end{equation}
If $f$ and $g$ coincide on $U=V(G)\backslash S$ then $f-g$ belongs
to  $L_{2}(S)$ and since $S$ is a $\Lambda$-set  we have
$$
\|f-g\|\leq \Lambda\left\|\mathcal{L}(f-g)\right\|.
$$
 Assume that $\omega<1/\Lambda$ and that $f$ is not identical to $g$. We have the
following inequalities
$$
\|f-g\|\leq \Lambda\|\mathcal{L}(f-g)\|\leq \Lambda
\omega\|f-g\|<\|f-g\|, \omega<1/\Lambda,
$$
which provide the desired contradiction  if $f-g$ is not identical
zero. It proves the Theorem.
\end{proof}

 As it was mentioned in the Introduction
one cannot expect that non-trivial uniqueness sets there exist for
functions from every Paley-Wiener subspace.  But it is reasonable
to expect that uniqueness sets exist for Paley-Wiener spaces
$PW_{\omega}(G)$ with relatively small $\omega>0$. Indeed, a
direct calculation shows that for any graph $G$ spaces
$PW_{\omega}(G)$ with
$$
0<\omega<\sqrt{1+\frac{1}{d(G)}}=\Omega_{G}>1, d(G)=\max_{v\in
V(G)}d(v),
$$
have non-trivial uniqueness sets.

Here are two examples of Paley-Wiener spaces on graphs and their
uniqueness sets.

\bigskip

1. \textbf{Finite graphs}. If a set of vertices $V(G)$ of a graph
$G$ is finite then the spectrum of the Laplace operator is
discrete and the space $PW_{\omega}(G)$ is a span of
eigenfunctions whose eigenvalues $\leq \omega$. In this case if
$U$ is a uniqueness sets for a space $PW_{\omega}(G)$ then $|U|$
is at least a number of eigenvalues (with multiplicities) of
$\mathcal{L}$ on the interval  $[0,\omega]$.

\bigskip

2. \textbf{Lattice $\mathbb{Z}^{n}$}. We consider a
one-dimensional lattice $\mathbb{Z}$.  In this case there is  a
version of the Fourier transform $\mathcal{F}$ on the space
$L_{2}(\mathbb{Z})$ which  is defined by the formula
$$
\mathcal{F}(f)(\xi)=\sum_{k\in \mathbb{Z}}f(k)e^{ik\xi}, f\in
L_{2}(\mathbb{Z}), \xi\in [-\pi, \pi).
$$
It gives a unitary operator from $L_{2}(G)$ on the space
$L_{2}(\mathbb{T})=L_{2}(\mathbb{T}, d\xi/2\pi),$ where
$\mathbb{T}$ is the one-dimensional torus and $d\xi/2\pi$ is the
normalized measure.  One can verify the following formula
$$
\mathcal{F}(\mathcal{L_{\mathbb{Z}}}f)(\xi)=2\sin^{2}\frac{\xi}{2}\mathcal{F}(f)(\xi),
$$
where $\mathcal{L_{\mathbb{Z}}}$ is the Laplace operator on the
graph $\mathbb{Z}$. The next result is obvious.
\begin{thm} The spectrum of the Laplace operator $\mathcal{L}_{\mathbb{Z}}$
 on the one-dimensional
lattice $\mathbb{Z}$ is the set $[0, 2]$. A function $f$ belongs
to the space $PW_{\omega}(\mathbb{Z}), 0<\omega<2,$ if and only if
the support of $\mathcal{F}f$ is a subset $\Omega_{\omega}$ of
$[-\pi,\pi)$ on which $2\sin^{2}\frac{\xi}{2}\leq \omega$.

\end{thm}

 Our
nearest goal is to show that for a one-dimensional line graph
$\mathbb{Z}$  the estimates in Poincare inequalities of finite
 successive  sets of vertices can be computed explicitly.

Consider a set of successive vertices $S=\{v_{1},
v_{2},...,v_{N}\}\subset \mathbb{Z}, $ and the corresponding space
$L_{2}(S)$. If $bS=\{v_{0},v_{N+1}\}$ is the boundary of $S$, then
for any $\varphi\in L_{2}(S)$ the function
$\mathcal{L}_{\mathbb{Z}}\varphi$ has support on $S\cup b S$ and
$$
\mathcal{L}_{\mathbb{Z}}\varphi(v_{0})=-\varphi(v_{1}),
\mathcal{L}_{\mathbb{Z}}\varphi(v_{1})=2\varphi(v_{1})-\varphi(v_{2}),
$$
$$
\mathcal{L}_{\mathbb{Z}}\varphi(v_{N})=2\varphi(v_{N})-\varphi(v_{N-1}),
\mathcal{L}_{\mathbb{Z}}\varphi(v_{N+1})=-\varphi(v_{N}),
$$
and for any other $v_{j}$ with $2\leq j\leq N-1,$
$$
\mathcal{L}_{\mathbb{Z}}\varphi(v_{j})=-\varphi(v_{N-1})+2\varphi(v_{j})-\varphi(v_{N+1}).
$$
 Let $C_{2N+2}=\Gamma(S)$ be a cycle graph
$$
C_{2N+2}=\{u_{-N-1},u_{-N},...,u_{-1}, u_{0},
u_{1},u_{2},...,u_{N}, u_{N+1}\}
$$
with  the following identification
$$
u_{-N-1}= u_{N+1}.
$$
Thus the total number of vertices in $C_{2N+2}$ is $2N+2$. We
introduce an embedding of $S\cup bS$ into $C_{2N+2}$ by the
following identification
$$v_{_{0}}=u_{0},
v_{1}=u_{1},...,v_{N}=u_{N},v_{N+1}=u_{N+1}.
$$
This embedding gives a rise to an embedding of $L_{2}(S)$ into
$L_{2}(C_{2N+2})$, namely every $\varphi\in L_{2}(S)$ is
identified with a function $F_{\varphi}\in L_{2}(C_{2N+2})$ for
which
$$
F_{\varphi}(u_{0})=0, F_{f\varphi}(u_{1})=\varphi(v_{1}), ...,
F_{\varphi}(u_{N})=\varphi(v_{N}), F_{\varphi}(u_{N+1})=0,
$$
and also
$$
 F_{\varphi}(u_{-1})=-\varphi(v_{1}), ..., F_{\varphi}(u_{-N})=-\varphi(v_{N}).
$$
It is important to note  that
$$
\sum_{u\in C_{2N+2}}F_{\varphi}(u)=0.
$$
If  $\mathcal{L}_{C}$ is the Laplace operator on the cycle
$C_{2N+2}$ then a direct computation  shows that for the vector
$F_{\varphi}$ defined above  the following is true
$$
2\|\varphi\|=\|F_{\varphi}\|,
2\|\mathcal{L}_{\mathbb{Z}}\varphi\|=\|\mathcal{L}_{C}F_{\varphi}\|,\varphi\in
L_{2}(S),F_{\varphi}\in L_{2}(C_{2N+2}).
$$
The operator $\mathcal{L}_{C}$ in $ L_{2}(C_{2N+2})$ has a
complete system of orthonormal eigenfunctions
\begin{equation}
\psi_{n}(k) =\exp 2\pi i\frac{n}{2N+2}k, 0\leq n\leq 2N+1,1\leq
k\leq 2N+2,
\end{equation}
with eigenvalues
\begin{equation}
\lambda_{n}=1-\cos \frac{2\pi n}{2N+2}, 0\leq n\leq 2N+1.
\end{equation}
The definition of  the function $F_{\varphi}\in L_{2}(C_{2N+2})$
implies that it  is orthogonal to all constants and its Fourier
series does not contain a term which corresponds to the index
$n=0$. It allows to obtain the following estimate
$$
\|\mathcal{L}_{C}F_{\varphi}\|^{2}=\sum
_{n=1}^{2N+1}\lambda_{n}^{2}\left|\left<F_{\varphi},\psi_{n}\right>\right|^{2}\geq
4\sin^{4}\frac{\pi}{2N+2}\|F_{\varphi}\|^{2}.
$$

It gives the following  estimate for functions $\varphi$ from
$L_{2}(S)$
$$
\|\varphi\|\leq \frac{1}{2}\sin^{-2}\frac {\pi}{2N+2}
\|\mathcal{L}_{\mathbb{Z}}\varphi\|.
$$

Thus we  proved the following Lemma.

\begin{lem}
If $S=\{v_{1},v_{2},...,v_{N}\}$ consists of $|S|=N$ successive
vertices of a line graph $\mathbb{Z}$ then it is a $\Lambda$-set
for
$$
\Lambda=\frac{1}{2}\sin^{-2}\frac{\pi}{2|S|+2}.
$$
In other words, for any $\varphi\in L_{2}(S)$ the following
inequality holds true
$$
\|\varphi\|\leq \Lambda\|\mathcal{L}_{\mathbb{Z}}\varphi\|.
$$
\end{lem}

\begin{rem} The last inequality which can be written as
$$
\|\mathcal{L}_{\mathbb{Z}}\varphi\|\geq
 2\sin^{2}\frac{\pi}{2|S|+2}\|\varphi\|, \varphi\in L_{2}(S),
$$
is similar to one of inequalities in \cite{FTT}.
\end{rem}
Note that in the case $|S|=1$ the last Lemma gives the inequality
$$
\|\delta_{v}\|\leq \|\mathcal{L}_{\mathbb{Z}}\delta_{v}\|,
S=\{v\},
$$
but direct calculations give a better value for $\lambda$:
$$
\|\delta_{v}\|=\sqrt{\frac{2}{3}}\left\|\mathcal{L}_{\mathbb{Z}}\delta_{v}\right\|,
v\in V.
$$

Let us note that if $\{S_{j}\}$ is a finite or infinite sequence
of disjoint subsets of vertices  $S_{j}\subset V$ such that the
sets $S_{j}\cup bS_{j}$ are pairwise disjoint and every $S_{j}$
has type $\Lambda_{j}$, then their union $S=\bigcup_{j} S_{j}$ is
a set of type $\Lambda=\sup_{j} \Lambda_{j}$. Indeed,  since the
sets $S_{j}$ are disjoint every function $\varphi\in L_{2}(S),
S=\bigcup_{j} S_{j}$, is a sum of functions $\varphi_{j}\in
L_{2}(S_{j})$ which are pairwise  orthogonal. Moreover because the
sets $S_{j}\cup bS_{j}$ are disjoint the functions
$\mathcal{L}\varphi_{j}$ are also orthogonal. Thus we have
$$
\|\varphi\|^{2}=\sum_{j}\|\varphi_{j}\|^{2}\leq
\sum_{j}\Lambda_{j}^{2}\|\mathcal{L}\varphi_{j}\|^{2}\leq
\Lambda^{2}\|\mathcal{L}\varphi\|^{2},
$$
where $\Lambda=\sup_{j}\Lambda_{j}$.

 A combination of this observation along with the last Lemma 3.6
  gives the following result for any $0<\omega<\sqrt{3/2}$.

\begin{thm}
If  $S$ is a finite or infinite union of
 disjoint sets $\{S_{j}\}$ of successive vertices such
that

1) the sets $\overline{S}_{j}=S_{j}\cup b S_{j}$ are disjoint

and

2) for every $j$ the following inequality holds
\begin{equation}
|S_{j}|< \frac{\pi}{2\arcsin\sqrt{\frac{ \omega}{2}}}-1,
\end{equation}
then every function $f\in PW_{\omega}(\mathbb{Z})$ is uniquely
determined by its values on the set $U=V(\mathbb{Z})\backslash S$.
\end{thm}

\bigskip

 A similar result holds true for a lattice
$\mathbb{Z}^{n}$ of any dimension. Consider for example the case
$n=2$. In this situation  the Fourier transform $\mathcal{F}$ on
the space $L_{2}(\mathbb{Z}^{2})$ is the  unitary operator
$\mathcal{F}$ which is defined by the formula
$$
\mathcal{F}(f)(\xi_{1}, \xi_{2})=\sum_{(k_{1},k_{2})\in
\mathbb{Z}^{2}}f(k_{1}, k_{2})e^{i k_{1}\xi_{1}+i k_{2}\xi_{2}},
f\in L_{2}(\mathbb{Z}\times \mathbb{Z}),
$$
where $ (\xi_{1},\xi_{2})\in [-\pi, \pi)\times [-\pi, \pi)$. The
operator $\mathcal{F}$ is isomorphism  of the space $L_{2}(G)$ on
the space $L_{2}(\mathbb{T}\times
\mathbb{T})=L_{2}(\mathbb{T}\times \mathbb{T},
d\xi_{1}d\xi_{2}/4\pi^{2}),$ where $\mathbb{T}$ is the
one-dimensional torus. the following formula holds true
$$
\mathcal{F}(\mathcal{L}_{\mathbb{Z}^{2}}f)(\xi)=\left(\sin^{2}\frac{\xi_{1}}{2}+
\sin^{2}\frac{\xi_{2}}{2}\right)\mathcal{F}(f)(\xi),
$$
where $\mathcal{L}_{\mathbb{Z}^{2}}$ is the Laplace operator on
the graph $\mathbb{Z}^{2}$. We have the following result.
\begin{thm} The spectrum of the Laplace operator on the lattice $\mathbb{Z}^{2}$
 is the set $[0,2]$.  A function $f$ belongs to the space
$PW_{\omega}(\mathbb{Z}\times \mathbb{Z}), 0<\omega<2,$ if and
only if the support of $\mathcal{F}f$ is a subset
$\Omega_{\omega}$ of $[-\pi,\pi)\times [-\pi,\pi)$ on which
$$
\sin^{2}\frac{\xi_{1}}{2}+ \sin^{2}\frac{\xi_{2}}{2}\leq \omega.
$$
\end{thm}
 Given a set $S=\{v_{n,m}\}, 1\leq n\leq N, 1\leq
m\leq M,$ we consider embedding of $S$ into two-dimensional
discrete torus of the size $T=(2N+2)\times (2M+2)=\{u_{n,m}\}$.
Every $f\in L_{2}(S)$ is identified with a function $g\in
L_{2}(T)$ in the following way
$$
g(u_{n,m})=f(v_{n,m}),1\leq n\leq N, 1\leq m\leq M,
$$
and
$$
g(u_{n,m})=0, N<n\leq N+2, M<m\leq M+2.
$$

We have
$$
\|\mathcal{L}_{\mathbb{Z}^{2}}f\|=\|\mathcal{L}_{T}g\|
$$
where $\mathcal{L}_{T}$ is the combinatorial Laplacian on the
discrete torus $T$. Since eigenfunctions of $\mathcal{L}_{T}$ are
products of the corresponding functions (3.1) a  direct
calculation gives the following inequality
$$
\|\varphi\|\leq \frac{1}{4}\frac{1}{\min\left(\sin
\frac{\pi}{2N+2},\sin
\frac{\pi}{2M+2}\right)}\|\mathcal{L}_{\mathbb{Z}^{2}}\varphi\|,
\varphi\in L_{2}(S).
$$

In a similar way one can  obtain corresponding results for a
lattice $\mathbb{Z}^{n}$ of any dimension. Note that the spectrum
of the Laplace operator on $\mathbb{Z}^{n}$ is $[0, 2]$ and
$\Omega_{\mathbb{Z}^{n}}=\sqrt{(2n+1)/2n}$.

Let $N_{j}=\{N_{1,j},...,N_{n,j}\}, j\in \mathbb{N}, $  be a
sequence $n$-tuples of natural numbers.  For every $j$ the
notation $S(N_{j})$ will be used for a "rectangular solid" of
"dimensions" $N_{1,j}\times N_{2,j}\times...\times N_{n,j}$.

Using these notations we  formulate the following sampling
Theorem.

\begin{thm}

If $S$ is a finite or infinite union of rectangular solids
$\{S(N_{j})\}$
    of vertices of dimensions  $N_{1,j}\times N_{2,j}\times...\times N_{n,j}$
    such that

    1) the sets $\overline{S}_{j}=S(N_{j})\cup b S(N_{j})$ are disjoint,

and

    2) the following inequality holds true for all $j$
$$
\omega< 4\min\left(\sin \frac{\pi}{2N_{1,j}+2},\sin
\frac{\pi}{2N_{2,j}+2},...,\sin \frac{\pi}{2N_{n,j}+2}\right),
$$
then every $f\in PW_{\omega}(\mathbb{Z}^{n})$ is uniquely
determined by its values on $U=V(\mathbb{Z}^{n})\backslash S$.

\end{thm}

\section{Reconstruction of Paley-Wiener spaces using splines}

 Now we are going to use variational splines
$Y_{k,\varepsilon}^{U, f}$ as a reconstruction tool of
Paley-Wiener functions $f\in PW_{\omega}(G)$ from their values on
uniqueness sets of the form $U=V(G)\setminus S$, where $S$ is a
$\Lambda$-set and $\Lambda<1/\omega$. We will need the following
Lemma.

\begin{lem}
If $A$ is a bounded self-adjoint positive definite operator in a
Hilbert space $H$ and for an  $\varphi\in H$ and a positive $a>$
the following inequality holds true
$$
\|\varphi\|\leq a\|A\varphi\|,
$$ then for the same $\varphi \in H$, and all $ k=2^{l}, l=0,1,2,...$ the following
inequality holds
$$
\|\varphi\|\leq a^{k}\|A^{k}\varphi\|.
$$
\end{lem}

\begin{proof}
By the spectral theory \cite{BS} there exist a direct integral of
Hilbert spaces
$$
X=\int_{0}^{\|A\|} X(\tau)dm (\tau )
$$
 and a unitary operator
$F$ from $H$ onto $X$, which transforms domain of $A^{t}, t\geq
0,$ onto $X_{t}=\{x\in X|\tau^{t}x\in X \}$ with norm
$$
\|A^{t}f\|_{H}=\left (\int_{0}^{\|A\|} \tau^{2t} \|Ff(\tau
)\|^{2}_{X(\tau)} d m(\tau) \right )^{1/2}
$$
and $F(A^{t} f)=\tau ^{t} (Ff)$. According to our assumption we
have for a particular  $\varphi\in H$
$$
\int _{0}^{\|A\|}| F\varphi(\tau)|^{2}d m(\tau)\leq a^{2} \int
_{0}^{\|A\|}\tau ^{2}| F\varphi(\tau)|^{2}d m(\tau)
$$
and then for the interval $B=B(0, a^{-1})$ we have
$$\int _{B}| F\varphi(\tau)|^{2}d m(\tau)+
\int_{[0, \|A\|]\setminus B}|F\varphi|^{2}d m(\tau)\leq
$$
$$
a^{2}\left( \int_{B}\tau^{2}|F\varphi|^{2}d m(\tau) +\int_{[0,
\|A\|]\setminus B}\tau^{2}|F\varphi| ^{2}d m(\tau)\right ) .
$$
 Since $a^{2}\tau^{2}<1$ on $B(0, a^{-1})$
$$
0\leq \int_{B}\left
(|F\varphi|^{2}-a^{2}\tau^{2}|F\varphi|^{2}\right)d m(\tau) \leq
\int _{[0, \|A\|]\setminus B}\left ( a^{2}
\tau^{2}|F\varphi|^{2}-|F\varphi|^{2}\right)d m(\tau).
$$
This inequality
 implies the inequality
 $$
 0\leq \int_{B}\left (a^{2}\tau^{2}|F\varphi|^{2}-
a^{4}\tau^{4}|F\varphi|^{2}\right)d m(\tau) \leq \int_{[0,
\|A\|]\setminus B}\left ( a^{4}
\tau^{4}|F\varphi|^{2}-a^{2}\tau^{2}|F\varphi| ^{2}\right)d
m(\tau)$$
 or
  $$a^{2}\int_{[0, \|A\|]}\tau^{2}|F\varphi|^{2}d m(\tau) \leq
 a^{4}\int_{\mathbb{R}_{+}}\tau^{4}|F\varphi|
 ^{2}d m(\tau),
 $$
 which means
 $$
 \|\varphi\|\leq a\|A\varphi\|\leq a^{2}\|A^{2}\varphi\|.
 $$
 Now, by using induction one can finish the proof of the
Lemma. The Lemma is proved.
\end{proof}
\textbf{Proof of the first part of the Theorem 1.1.}

We assume that the operator $\mathcal{L}$ has bounded inverse and
$\varepsilon=0$.
 If $f\in PW_{\omega}(G)$ and
$Y_{k}^{U, f},\left( Y_{k}^{U, f}=Y_{k,0}^{U, f}\right)$ is a
variational spline which interpolates $f$ on  a set
$U=V(G)\setminus S$, where $S$  is a $\Lambda$- set
$\left(0<\omega<1/\Lambda\right)$, then
 $f-Y_{k}^{U, f}\in L_{2}(S)$ and we have
\begin{equation}
\|f-Y_{k}^{U, f}\|\leq \Lambda\|\mathcal{L}(f-Y_{k}^{U, f})\|.
\end{equation}

At this point we can apply the last Lemma with $A=\mathcal{L}$,
$a=\Lambda$ and $\varphi=f-Y_{k}^{U, f}$. It gives the inequality
\begin{equation}
\|f-Y_{k}^{U, f}\|\leq \Lambda^{k}\|\mathcal{L}^{k}(f-Y_{k}^{U,
f})\|
\end{equation}
for all $ k=2^{l}, l=0,1,2,...$ Since the interpolant $Y_{k}^{U,
f}$ minimizes the norm $\|\mathcal{L}^{k}\cdot\|$ it gives
$$
\|f-Y_{k}^{U, f}\|\leq 2\Lambda^{k}\|\mathcal{L}^{k}f\|, k=2^{l},
l\in \mathbb{N}.
$$
Because for functions  $f\in PW_{\omega}(G)$ the Bernstein
inequality  holds
$$
\|\mathcal{L}^{m}f\|\leq \omega^{m}\|f\|, m\in \mathbb{N},
$$
it implies the first part of the Theorem
1.1:
$$
\|f-Y_{k}^{U, f}\|\leq 2\gamma^{k}\|f\|, \gamma=\Lambda\omega<1,
k=2^{l}, l\in \mathbb{N}.
$$

\textbf{Proof of the second part of the Theorem 1.1.}

Now we assume that the operator $\mathcal{L}$ is not invertible
(it is a typical situation on any finite graph). We fix an
$$
0<\varepsilon <\frac{1}{\Lambda}
$$ and assume that
$$
0<\omega<\frac{1}{\Lambda}-\varepsilon.
$$
 If $f\in PW_{\omega}(G)$ and
$Y_{k,\varepsilon}^{U, f}$ is a variational spline which
interpolates $f$ on a set $U=V(G)\setminus S$ where $S$  is a
$\Lambda$- set then
 $f-Y_{k,\varepsilon}^{U, f}\in L_{2}(S)$ and we have
\begin{equation}
\|f-Y_{k,\varepsilon}^{U, f}\|\leq
\Lambda\|\mathcal{L}(f-Y_{k,\varepsilon}^{U, f})\|.
\end{equation}

For any $g\in L_{2}(G)$ the following inequality holds true
\begin{equation}
\|\mathcal{L}g\|\leq \|(\varepsilon I+\mathcal{L})g\|.
\end{equation}

Thus the inequalities (4.3) and (4.4)  imply the inequality
$$
\|f-Y_{k,\varepsilon}^{U, f}\|\leq \Lambda\|(\varepsilon
I+\mathcal{L})(f-Y_{k,\varepsilon}^{U, f})\|.
$$

We apply the  Lemma 4.1 with $A=\varepsilon I+\mathcal{L}$,
$a=\Lambda$ and $\varphi=f-Y_{k,\varepsilon}^{U, f}$. It gives the
inequality
$$
\|f-Y_{k,\varepsilon}^{U, f}\|\leq \Lambda^{k}\|(\varepsilon
I+\mathcal{L})^{k}(f-Y_{k,\varepsilon}^{U, f})\|
$$
for all $ k=2^{l}, l=0,1,2,...$
 Using the minimization property of $Y_{k,\varepsilon}^{U, f}$ we
obtain
$$
\|f-Y_{k,\varepsilon}^{U, f}\|\leq 2\Lambda^{k}\|(\varepsilon
I+\mathcal{L})^{k}f\|,k=2^{l}, l\in \mathbb{N}.
$$
If $f\in PW_{\omega}(G)$, then the Bernstein inequality
$$
\|\mathcal{L}^{m}f\|\leq \omega^{m}\|f\|, m\in \mathbb{N},
$$
implies  the inequality
$$
\|(\varepsilon I+\mathcal{L})^{m}f\|\leq
(\omega+\varepsilon)^{m}\|f\|,m\in \mathbb{N}.
$$
 After all we have the following
inequality
$$
\|f-Y_{k,\varepsilon}^{U, f}\|\leq 2\gamma^{k}\|f\|,
\gamma=\Lambda(\omega+\varepsilon)<1,k=2^{l}, l\in \mathbb{N}.
$$
The proof of the Theorem 1.1 is complete.

\end{document}